\documentclass[11pt]{article}
\pdfoutput=1

\usepackage{cite}
\usepackage{helvet}
\usepackage{amsmath}
\usepackage{amssymb}
\usepackage{setspace}
\usepackage{graphicx}
\usepackage{empheq}
\usepackage{soul}
\usepackage{subcaption}
\usepackage{tabularx}
\usepackage{array}
\usepackage{float}
\usepackage{braket}
\usepackage[utf8]{inputenc}
\usepackage{url}
\usepackage{hyperref}
\usepackage{slashed}

\def\be{\begin{equation}}
\def\ee{\end{equation}}

\newcommand{\MSbar}{\overline{\text{MS}}}

\usepackage[a4paper,top=3cm,bottom=3cm,left=2.5cm,right=2.5cm,bindingoffset=0mm]{geometry}

\begin{document}

\begin{center}
{\Large \bf The Proton in High Definition: Revisiting \\ \vspace{4mm} Photon--Initiated Production in High Energy Collisions}

\vspace*{1cm}
                                                   
L. A. Harland--Lang \\                                                 
                                                   
\vspace*{0.5cm}
${}^1$Rudolf Peierls Centre, Beecroft Building, Parks Road, Oxford, OX1 3PU                                          
                                                    
\vspace*{1cm}

\begin{abstract}
\noindent
We re--examine the current state of the art for the calculation of photon--initiated processes at the LHC, as formulated in terms of a photon PDF in the proton that may be determined rather precisely from the known proton structure functions. We in particular demonstrate that a by construction more precise calculation is provided by a  direct application of the structure function approach, best known from the case of Higgs Boson production via vector boson fusion. This avoids any artificial scale variation uncertainties, which can otherwise be rather significant for processes calculated within the standard approach thus far. To understand the source of these, we present a detailed comparison of the structure function approach and its relation to the photon PDF. We then provide precise predictions for the photon--initiated contribution to lepton pair production at the LHC, including the lepton pair transverse momentum distribution. Thus, by a direct application of the structure function formalism we show how the contribution from initial--state photons at the LHC may for the first time be included with high precision in a universal and straightforward way, providing a high definition picture of the photon content of the proton.
\end{abstract}

\end{center}

\section{Introduction}

A full and precise account of photon--initiated contributions to LHC processes has become mandatory in light of the high precision standard being aimed for at the LHC, both in terms of the theoretical inputs and the experimental data itself. In light of this the study of photon--initiated production at the LHC has undergone significant progress in recent years. Such studies have all been based on the idea of explicitly including the photon as an additional partonic constituent of the proton, which mandates the introduction of a photon PDF within the proton. The photon--initiated cross section can then be calculated using the standard framework of collinear factorization, in much the same way as the usual quark/gluon--induced QCD processes.

Within this framework, the basic aim has been to achieve a precise determination of the photon PDF itself. Earlier work focussed on the calculation of this within phenomenological models~\cite{Martin:2004dh,Schmidt:2015zda} of photon emission from the quarks within the proton, or on completely agnostic fits ~\cite{Ball:2013hta,Giuli:2017oii} to Drell--Yan data. The importance of elastic photon emission was emphasised in~\cite{Harland-Lang:2016apc,Harland-Lang:2016kog}, which provides an important component of the photon PDF and is theoretically well understood in terms of the elastic structure functions of the proton. Indeed, this idea is rather an old one, due to the so--called equivalent photon approximation (EPA)~\cite{Budnev:1974de}, within which the photon PDF corresponds to the photon flux emitted by the proton, with contributions from both elastic and inelastic emission that are directly related to the the corresponding structure functions ($F_{1,2}^{\rm el}$, $F_{1,2}^{\rm inel.}$) of the proton (see also~\cite{Anlauf:1991wr,Blumlein:1993ef,Mukherjee:2003yh,Luszczak:2015aoa} for earlier work). 

This idea was put on a precise theoretical and phenomenological footing by the LUXqed group~\cite{Manohar:2016nzj,Manohar:2017eqh}, who both demonstrated how the concept of an EPA flux could be extended beyond LO within the collinear factorization framework, combining consistently with higher order quark/gluon--initiated diagrams, and provided the first serious phenomenological input for the required structure functions to give the first publicly available PDF set within such an approach. Following from this, photon PDFs have been provided in combination with the global MMHT~\cite{Harland-Lang:2019pla} and NNPDF~\cite{Bertone:2017bme} sets, in both cases closely following the same approach as LUXqed. In all cases, the experimental input for the corresponding structure function inputs is sufficiently precise that the quoted photon PDF uncertainty is very small, generally at the $\sim 1 \%$ level, and thus a high precision photon PDF determination can be claimed.

However, while the collinear photon PDF as defined in the above procedure is indeed known to this high level of precision, this is only one ingredient in the calculation of photon--initiated production at the LHC. In particular, one also requires the corresponding parton--level cross sections as input, for which only the leading order contributions are often included in phenomenological studies. Indeed, currently only a rather limited set of studies including NLO EW corrections to photon--initiated production, that is via $q \to q \gamma$ splittings in the initial state, are available~\cite{Dittmaier:2009cr,Baglio:2013toa,Kallweit:2017khh}. As we will discuss, this introduces a source of uncertainty in the predicted photon--initiated cross section that is often significantly larger than that implied by the high precision determination of the photon PDF itself.

In this work, we will show how a straightforward calculation of photon--initiated production is provided by simply applying the well known `structure function' approach~\cite{Han:1992hr}, which has for many years formed the central element in the calculation of Higgs boson production via vector boson fusion (VBF), at NLO~\cite{Figy:2003nv}, NNLO~\cite{Bolzoni:2010xr,Cacciari:2015jma,Cruz-Martinez:2018rod} and N${}^3$LO in QCD~\cite{Dreyer:2016oyx}. This bypasses any explicit reference to the photon PDF, and provides predictions which are by construction more precise than any of the currently available calculations within the standard collinear factorization approach. Percent level precision in the predicted cross sections is in particular achieved for the first time, with remaining contributions from e.g. the `non--factorizable' corrections that break the structure function picture expected to be small, as they are in the VBF Higgs case.

We present a detailed discussion of this approach, and its connection to the photon PDF formalism, via the equivalent photon approach and its extension in the LUXqed framework. The key point is that the current PDF approach uses exactly the same input as the structure function calculation, but in an approximate form. We in particular examine within the simpler context of lepton--hadron (and photon--hadron) scattering the degree of discrepancy and uncertainty that is introduced by applying the collinear factorization framework, and which is absent in the structure function approach. At LO in $\alpha$ these are found to be very large, significantly so in comparison to the quoted photon PDF uncertainties, with the scale variation band often not overlapping at all with the full result. At NLO the situation is naturally found to improve, but nonetheless the uncertainty is often comparable to or larger than the PDF uncertainty. When more exclusive observables are considered, such as the differential cross section with respect to the photon $Q^2$, this discrepancy is larger still. We in addition discuss the connection to the $k_\perp$--factorization approach (see e.g.~\cite{daSilveira:2014jla,Luszczak:2015aoa,Dyndal:2019ylt} for applications in this context), and demonstrate the deficiencies of this in comparison to the full calculation. This discussion sheds light in particular on the differences between the collinear and $k_\perp$--factorization predictions seen in e.g.~\cite{Dyndal:2019ylt}.

We then move to the case of proton--proton collisions at the LHC, focussing on the case of lepton pair production. We compare the structure function calculation and predictions within collinear factorization for lepton pair production in the kinematic regime relevant to the LHC, from low to high masses. We find that the latter have large uncertainties associated with them that are straightforwardly bypassed by applying the structure function approach. Finally we present predictions for the photon--initiated contribution to the lepton pair transverse momentum distribution.  Within the collinear calculation this only begins at NLO in $\alpha$, with the leading $q\gamma$--initiated contribution presented some time ago in~\cite{Arbuzov:2007kp,CarloniCalame:2007cd}. Our results represent the first high precision prediction for this observable, which for the first time are applicable from zero to high $p_\perp^{ll}$, that is in the kinematic regions relevant to both the fixed order pQCD calculation of the Drell--Yan process and the region of lower transverse momenta, where resummation must be applied. We in particular present results for the ATLAS 8 TeV event selection~\cite{Aad:2015auj}. These can enter at the level of a few percent in the region where fixed--order QCD may be applied, relevant to PDF fits, while for the lower $p_\perp^{ll}$ region relevant to comparisons with resummed QCD calculations, these can be as large as 10\%. 

The outline of this paper is as follows. In Section~\ref{sec:structfun} we summarise the key ingredients of the structure function approach. In Section~\ref{sec:lephad} we present a detailed comparison of this with the standard approach, in terms of a photon PDF, for the simpler case of lepton--proton (and photon--proton) scattering; we in particular demonstrate explicitly how the standard approach is derived via an approximation to the structure function calculation, and which is therefore by construction less precise. In Section~\ref{sec:hadhad} we discuss the case of proton--proton collisions, and present phenomenological predictions for lepton pair production at the LHC. In Section~\ref{sec:conc} we conclude and discuss future work.

\section{Structure Function Calculation}\label{sec:structfun}

The basic observation we apply is that in the high--energy limit the photon--initiated cross section in proton--proton collisions\footnote{We will for concreteness consider the case of two--photon initiated production, but the mixed case where only one photon participates in the initial state can be written down in a similar way.} can be written in the general form
  \be\label{eq:sighh}
  \sigma_{pp} = \frac{1}{2s} \int \frac{{\rm d}^3 p_1' {\rm d}^3 p_2' {\rm d}\Gamma}{E_1' E_2'}   \alpha(Q_1^2)\alpha(Q_2^2)
  \frac{\rho_1^{\mu\mu'}\rho_2^{\nu\nu'} M^*_{\mu'\nu'}M_{\mu\nu}}{q_1^2q_2^2}\delta^{(4)}(q_1+q_2 - k)\;.
 \ee
 Here the outgoing hadronic systems have momenta $p_{1,2}'$ and the photons have momenta $q_{1,2}$, with $q_{1,2}^2 = -Q_{1,2}^2$. We consider the production of a system of 4--momentum $k = q_1 + q_2 = \sum_{j=1}^N k_j$ of $N$ particles, where ${\rm d}\Gamma = \prod_{j=1}^N {\rm d}^3 k_j / 2 E_j (2\pi)^3$ is the standard phase space volume. $M^{\mu\nu}$ corresponds to the $\gamma\gamma \to X(k)$ production amplitude, with arbitrary photon virtualities. 
 
 The above expression is the basis of the equivalent photon approximation~\cite{Budnev:1974de}, as well as being precisely the formulation used in the structure function approach~\cite{Han:1992hr} applied to the calculation of Higgs Boson production via VBF. In particular, $\rho$ is the density matrix of the virtual photon, which is given in terms of the well known proton structure functions:
 \be
 \rho_i^{\alpha\beta}=2\int \frac{{\rm d}M_i^2}{Q_i^2}  \bigg[-\left(g^{\alpha\beta}+\frac{q_i^\alpha q_i^\beta}{Q_i^2}\right) F_1(x_{B,i},Q_i^2)+ \frac{(2p_i^\alpha-\frac{q_i^\alpha}{x_{B,i}})(2p_i^\beta-\frac{q_i^\beta}{x_{B,i}})}{Q_i^2}\frac{ x_{B,i} }{2}F_2(x_{B,i},Q_i^2)\bigg]\;,
 \ee
where $x_{B,i} = Q^2_i/(Q_i^2 + M_{i}^2 - m_p^2)$ for a hadronic system of mass $M_i$ and we note that the definition of the photon momentum $q_i$ as outgoing from the hadronic vertex is opposite to the usual DIS convention. Here, the integral over $M_i^2$ is understood as being performed simultaneously with the phase space integral over $p_{i}'$, i.e. is not fully factorized from it (the energy $E_i'$ in particular depends on $M_i$)\footnote{We note that in the published version of this paper, the expression valid for fixed $Q^2$ is instead given, which is not consistent with \eqref{eq:sighh} as written. However the above formulation is used throughout.}. This corresponds to the general Lorentz--covariant expression that can be written down for the photon--hadron vertex, and indeed because of precisely this point it is the same object which appears in the cross section for (photon--initiated) lepton--hadron scattering, including in the DIS region. We have
\be\label{eq:siglp}
\frac{{\rm d}\sigma_{lp}}{{\rm d}Q^2} = \frac{\alpha(Q^2)}{4 s^2} \frac{\rho_i^{\alpha\beta}  L_{\alpha\beta}}{Q^2}\;,
\ee
where $L$ is the usual spin--averaged leptonic tensor. Indeed the photon density matrix is straightforwardly related to the standard hadronic tensor $W^{\alpha \beta}$ that enters the e.g. the DIS cross section via
\be
\rho_i^{\alpha\beta}= 2 \int \frac{{\rm d}M_i^2}{Q_i^2}\,W^{\alpha \beta}_i  =2\int \frac{{\rm d}x_{B,i}}{x_{B,i}^2}  \,W^{\alpha \beta}_i \;,
\ee
where the second relation holds at fixed $Q^2_i$. One can then as usual extract  $F_{1,2}$ from the measured cross sections for lepton--proton scattering. For the case of VBF, the procedure is precisely the same, but one instead considers the structure functions related to the weak current. In this way, our general expression \eqref{eq:sighh}, combined with a suitable input for the proton structure functions, represents the complete result we need to calculate the corresponding photon--initiated cross section in proton--proton collisions. In particular, no explicit reference is made to the partonic content of the proton itself, including the photon PDF. 

What are the uncertainties in the above approach? First, we have the more straightforward uncertainty due to the experimental determination of the structure functions, which enters in exactly the same way as for the extraction of the photon PDF~\cite{Manohar:2016nzj,Manohar:2017eqh,Bertone:2017bme,Harland-Lang:2019pla}. At low $Q^2$, we must include the experimental uncertainty in the publicly available fits to the elastic and inelastic (resonant and non--resonant) structure functions, allowing in principle for the uncertainty due to higher--twist/rernormalon corrections in $F_L$, which play some role. At high $Q^2$, where the structure functions are best defined using theoretical pQCD expressions in combination with PDFs extracted from global analyses, we must include the usual PDF uncertainty, propagated in the standard way. For further details we refer the reader to~\cite{Manohar:2016nzj,Manohar:2017eqh,Bertone:2017bme,Harland-Lang:2019pla} and to the discussion in Section~\ref{sec:hadhad}, but we note here that such uncertainties generally enter at the percent level. At high $Q^2$, there is in addition a small uncertainty due to the missing higher order corrections beyond the NNLO order in QCD at which we will work, though given the PDFs which enter these predictions are themselves to some extent fitted to structure function data in this $Q^2$ region, the interpretation of such an uncertainty is not completely clear, being somewhat overlapping with the experimental uncertainty on the structure functions themselves.

A more subtle question to the above relates to the so--called `non--factorizable' corrections. As in the case of Higgs production via VBF the structure function approach only includes the factorizable contributions to the cross section, that is it excludes all corrections due to QCD or EW exchanges between the protons. At parton--level and lowest non--trivial order this corresponds to gluon or electroweak boson exchange between the quark/antiquark lines undergoing the $q\to q\gamma$ splitting, or between these and the centrally produced final-state if it is permitted. However, as in the case of VBF the QCD corrections only occur at NNLO and are in addition colour suppressed, such that they are generally expected to be at the $\sim 0.5\%$ level for Higgs boson production via VBF~\cite{Liu:2019tuy}. The corresponding corrections for photon--initiated production processes should enter at a similar level, as should NLO EW corrections due to virtual photon/$Z$ exchange. The inclusion of such corrections would in general necessitate a departure from the structure function approach, though we note that in fact the calculation of~\cite{Liu:2019tuy}, for which the leading NNLO QCD corrections are directly proportional to the LO cross section, would in principle allow a rather straightforward inclusion of these into the structure function framework. However, as the size of these corrections are so small, and always smaller than the current uncertainty on the structure function inputs, they can be safely neglected. Moreover, as discussed in~\cite{Manohar:2017eqh} if one were to include such corrections it would also be necessary to include QED corrections to the structure functions, which would enter at the same level. This would call for a detailed assessment of the way in which QED corrections have been accounted for in published structure function data. Thus the inclusion of these corrections in either the structure function approach or indeed collinear factorization is not straightforward. Related to this, it is not necessarily useful to consider N${}^3$LO QCD corrections to the structure functions themselves, given such corrections should also be comparable to or smaller than these non--factorizable corrections. Finally, higher order corrections to the $\gamma\gamma\to X$ subprocess itself can be included in the usual way as in collinear factorization.

Now, to clarify some of the results which will follow, it is useful to consider first the case of lepton--proton scattering, or more generally any scattering process with a single proton in the initial state. We will in particular consider the relationship between the results of the structure function approach, the equivalent photon approximation and the collinear photon PDF. 

\section{Test Case: Lepton/Photon--hadron scattering}\label{sec:lephad}

We will consider the scattering of a (for simplicity) massless initial state $K(k)$ off a proton $P(p)$, producing a final state $K'(k')$ with mass $M$, and we will use the standard logic of the equivalent photon approximation~\cite{Budnev:1974de}. The general expression for this is given by \eqref{eq:siglp}, where the tensor $L$ depends on the particular $K$ and $K'$. We first expand this tensor $L^{\alpha\beta}$ as
\be\label{eq:Lexpan}
L^{\alpha\beta} = -\delta_T^{\alpha\beta} |L_T|^2 + \epsilon_{0}^\alpha\epsilon_{0}^\beta |L_0|^2\;,
\ee
where $L_T = L_\pm$ is the amplitude corresponding to $\pm$ photon helicities, and $L_0$ corresponds to longitudinal photon. Here
\be\label{eq:projdef}
\epsilon_{0}^\alpha = -\frac{\sqrt{Q^2}}{(k\cdot q)} \left( k^\alpha + \frac{q \cdot k}{Q^2}q^\alpha \right)\qquad \qquad
\delta_T^{\alpha\beta} =g^{\alpha\beta} + \frac{q^\alpha q^\beta}{Q^2} - \epsilon_{0}^\alpha\epsilon_{0}^\beta\;,
\ee
such that these project out the longitudinal and transverse photon helicities in the $\gamma^* K$ c.m.s. frame. We can write these in terms of the cross sections for transverse (`$T$') and longitudinal (`$0$') photon absorption via
\be
\sigma_{T,0} =-\frac{\pi}{2s (q\cdot k)} \delta\left(\xi- x\right)  |L_{T,0}|^2 \;,
\ee
where $x=M^2/s$ and $\xi=(k+q)^2/s$, which in the $Q^2 \ll M^2$ limit corresponds to the proton momentum fraction carried by the photon, and in what follows we assume that $s\gg m_p^2$ for simplicity. Note the overall sign is consistent with the fact that $q\cdot k$ is negative in our convention. Substituting this into \eqref{eq:siglp},  we have
\be\label{eq:sigepa1}
\sigma =  \int {\rm d} \xi\, \frac{\alpha}{2\pi} \int_{Q^2_{\rm min}}^{Q^2_{\rm max}} \frac{{\rm d} Q^2}{Q^2} \,x\frac{(q\cdot k)}{M^2}\, 
\left( \hat{\sigma}_T(\xi,Q^2)  \delta_T^{\alpha\beta} -\hat{\sigma}_0(\xi,Q^2)\epsilon_{0}^\alpha\epsilon_{0}^\beta \right)\rho_{\alpha\beta} \;.
\ee
 We then find that
\begin{align}\nonumber
 \delta_T^{\alpha\beta} \rho_{\alpha\beta} &=\int_x^1 \frac{{\rm d} z}{z} \frac{1}{x^2} \frac{M^2}{(q\cdot k)}  \bigg[\bigg( zp_{\gamma q}(z)+z^2\frac{Q^2}{M^2} + \frac{Q^2}{(q\cdot k)}+\frac{2m_p^2 x^2}{Q^2}\bigg) F_2(x/z,Q^2)\\ \label{eq:trandot} &-z^2\bigg(1+\frac{Q^2}{M^2} \bigg)F_L\!\left(\frac{x}{z},Q^2\right)\bigg]\;.
\end{align}
where $p_{\gamma q}$ is the usual LO splitting function, and
\be\label{eq:scaldot}
 \epsilon_{0}^\alpha\epsilon_{0}^\beta \rho_{\alpha\beta} =\int_x^1 \frac{{\rm d} z}{z} \frac{1}{x^2} \frac{M^2}{(q\cdot k)}  \bigg[\bigg(2z+\frac{M^2}{(q\cdot k)}\bigg) F_2(x/z,Q^2)-\frac{z^2}{2}\bigg(1+\frac{Q^2}{M^2} \bigg)F_L\!\left(\frac{x}{z},Q^2\right)\bigg]\;,
\ee
where $z=x/x_B$ and $F_L(x,Q^2) = (1+4m_p^2x^2/Q^2)F_2(x,Q^2) - 2x F_1(x,Q^2)$. Here we have assumed that $M^2 \gg m_p^2$ for simplicity, and will continue to do so in the discussion which follows, though in the numerics we keep everything exact. Note that under this assumption the kinematic limits on the $Q^2$ integral in \eqref{eq:sigepa1} are $Q_{\min}^2 = x^2 m_p^2/(1-z)$, $Q_{\max}^2 = M^2(1-z)/z$, while the limits on the $z$ integration also readily follow from the kinematic limits on $x_B$. The full form of the kinematic limits are given in~\cite{Manohar:2017eqh}, and are used in all numerics which follow.

So far, the above results are exact. To derive a simpler, approximate, expression we now apply the equivalent photon approximation. Namely, we consider the $M^2 \gg Q^2$ limit, for which the longitudinal cross section $\sigma_L \sim 0$ and the transverse cross section is independent of $Q^2$. We can therefore drop the contribution from \eqref{eq:scaldot}, while \eqref{eq:trandot} becomes
\be
\delta_T^{\alpha\beta} \rho_{\alpha\beta}\approx \int_x^1 \frac{{\rm d} z}{z} \frac{1}{x^2} \frac{M^2}{(q\cdot k)}   \bigg[\left( zp_{\gamma q}(z)+ \frac{2 x ^2 m_p^2}{Q^2}\right) F_2(x/z,Q^2)-z^2 F_L\!\left(\frac{x}{z},Q^2\right)\bigg]\;.
\ee
This allows us to write the simple result
\be\label{eq:sigepa2}
\sigma \approx  \int {\rm d} \xi\, \hat{\sigma}(\xi) \cdot f_{\gamma/p}^{\rm PF}(\xi,\mu^2)\;,
\ee
where we drop the `T' subscript, it being implicit that this is the transverse on--shell cross section for $\gamma K \to K'$, relevant to the $Q^2 \ll M^2$ limit, and 
\begin{align}\nonumber
  x f_{\gamma/p}^{\rm PF}(x,\mu^2) &= 
  \frac{1}{2\pi \alpha(\mu^2)} \!
  \int_x^1
  \frac{dz}{z}
  \int^{\frac{\mu^2}{1-z}}_{\frac{x^2 m_p^2}{1-z}} 
  \frac{dQ^2}{Q^2} \alpha^2(Q^2)
  \\  \label{eq:xfgamma-phys}
  &\cdot\Bigg[\!
  \left(
    zp_{\gamma q}(z)
    + \frac{2 x ^2 m_p^2}{Q^2}
  \right)\! F_2(x/z,Q^2)
    -z ^2
  F_L\!\left(\frac{x}{z},Q^2\right)
  \Bigg]\,,
\end{align}
which is precisely the `physical' photon PDF derived in~\cite{Manohar:2017eqh}. This therefore provides a simple formula for the photon--initiated cross section in terms of a photon PDF of the proton, and an on--shell photon--initiated production cross section. This however remains an approximation to the complete result given by direct application of \eqref{eq:siglp}, that is only valid in the $M^2 \gg Q^2$ limit. The complete result in particular accounts for the full $Q^2$ dependence of the hard matrix element, including the contribution from longitudinal photon polarizations. The reliability of this approximation will be examined below. 

We note that in the above expression we have introduced the `factorization' scale $\mu$. In the equivalent photon approximation no such new scale is directly introduced, and in principle one can evaluate the upper limit of the $Q^2$ integral by its proper kinematic endpoint. However, the sensitivity to this choice is beyond the $Q^2 \ll M^2$ approximation that has been made here, and so in essence could be viewed in the above case as a natural parameterisation of our sensitivity to this approximation, rather similar to the role that the factorization scale indeed plays in collinear factorization. The precise choice of $Q^2_{\rm max} \to \mu^2/(1-z)$ is made for consistency with the results below in the collinear framework, so that in the above case the choice of $\mu^2 = (1-z) Q^2_{\rm max}$ would be consistent with the kinematic upper limit. It is understood that the coupling $\alpha$ in the subprocess cross section $\hat{\sigma}$ should be evaluated at the scale $\mu^2$. We in particular recall, as discussed in~\cite{Harland-Lang:2016lhw,Kallweit:2017khh}, that the use of the on--shell renormalization scheme is not appropriate here; the coupling evaluated at $\mu^2$ will then cancel with the coupling in the denominator of \eqref{eq:xfgamma-phys}, leaving only the two powers of the coupling evaluated at the appropriate scale, namely the photon virtuality, $Q^2$.

To examine the connection between the systematic treatment of~\cite{Manohar:2016nzj,Manohar:2017eqh}   and the discussion above, we will consider the simple `toy' model considered in~\cite{Manohar:2017eqh}, namely the production via $l(k) + p(p) \to L(k') + X$ of a heavy lepton $L$. We will only highlight the points relevant to our discussion, while a detailed discussion of this toy model, and the derivation of the photon PDF can be found in~\cite{Manohar:2016nzj,Manohar:2017eqh}. The corresponding leptonic tensor reads $L^{\alpha\beta} = \frac{ \alpha c_0}{8\pi} {\rm Tr}\left(\slashed{k}[\slashed{q},\gamma^\alpha ](\slashed{k'}+M)[\gamma^\beta,\slashed{q}]\right)$, where $c_0$ is an overall constant. From this we find
\begin{align}
\hat{\sigma}_T(\xi,Q^2) &=\frac{\alpha c_0}{s}\delta\left(\xi -x\right) \cdot M^2\; ,\\
\hat{\sigma}_0(\xi,Q^2) &=\frac{\alpha c_0}{s}\delta\left(\xi -x\right) \cdot Q^2\;,
\end{align}
which indeed have the correct scaling behaviour at low $Q^2$ discussed above. Substituting these into \eqref{eq:sigepa1} we readily arrive at the result given in~\cite{Manohar:2017eqh}, which we reproduce here for clarity (again in the $M^2 \gg m_p^2$ limit for simplicity):
\begin{align}\nonumber
\sigma &= \frac{c_0}{2\pi} \int_x^{1} \frac{dz}{z}\int^{Q_\text{max}^2}_{Q_\text{min}^2} \frac{dQ^2}{Q^2} \alpha^2(Q^2)   \Bigg[
  \biggl(z p_{\gamma q}(z)  + \frac{2 x ^2 m_p^2}{Q^2}+\frac{z ^2 Q^2}{M^2} -\frac{2 z  Q^2}{M^2}\biggr) F_2(x/z,Q^2) \\ \label{eq:sigma-HL-simple}
  &+\left(
    -z ^2
    -\frac{z ^2 Q^2}{2 M^2}
    +\frac{z ^2 Q^4}{2 M^4}
  \right)F_L(x/z,Q^2)
  \Bigg]\;.
\end{align}
The corresponding expression for this given within the collinear factorization framework detailed in~\cite{Manohar:2016nzj,Manohar:2017eqh} is
\begin{equation}
  \label{eq:coll-fact-basic}
  \sigma = \sum_{a} \int_x^1 \frac{dz}{z}\, 
   \hat\sigma_{a}(z,\mu^2)\, 
  \frac{M^2}{zs} f_{a/p}\left(\frac{M^2}{zs},\mu^2\right)\,,
\end{equation}
with 
\be
  \label{eq:coll-sub-L-qe-alt}
  \hat \sigma_{a}(z,\mu^2) 
  = c_0 \,\alpha(\mu^2)\delta(1-z)\delta_{a\gamma} + c_0 \frac{\alpha^2(\mu^2)}{2\pi}\Bigg[ 
    f^{NL}(z) 
    + z p_{\gamma q}(z) \ln\frac{M^2(1-z)^2}{z\mu^2}
  \Bigg] \sum_{i\in \{q,\bar{q}\}} e_i^2 \delta_{a i}\,,
\ee
to $O(\alpha^2)$, in the $\MSbar$ factorization scheme, and where the non--logarithmic contribution $f^{NL}(z) = -2 + 3z$. The photon PDF is given by
\be\label{eq:xfgamma-MSbar}
 x f_{\gamma/p}(x,\mu^2) =  x f_{\gamma/p}^{\rm PF}(x,\mu^2) - \frac{\alpha(\mu^2)}{2\pi }   \int_x^1  \frac{dz}{z} z^2 F_2\left(\frac{x}{z},\mu^2\right)\;.
\ee
 Keeping just the explicit leading order in $\alpha$ contribution to the cross section \eqref{eq:coll-sub-L-qe-alt} and dropping the higher order in $\alpha$ matching term in the photon PDF, we can see that this is consistent with the result of \eqref{eq:sigepa2}, as expected. In particular, the overall normalization of $\hat{\sigma}$ is consistent with the interpretation as the cross section for on--shell photon absorption as in $\hat{\sigma}_T$. Beyond LO however, the relationship between this and $\hat{\sigma}_{L,T}$ is not as straightforward.

The role of the $O(\alpha^2)$ quark--initiated contribution to the cross section and the matching term in the photon PDF (that is, the second term in \eqref{eq:coll-sub-L-qe-alt}) is discussed in detail in~\cite{Manohar:2017eqh}. Here we simply recall what these achieve from a practical point of view. In particular, these operate in the 
\be
\sum_{i\in \{q,\bar{q}\}}  x f_i(x,\mu^2) \approx F_2(x,\mu^2) \approx F_2(x,Q^2)\;,
\ee
approximation, which is true up to $O(\alpha_S, \alpha)$ corrections. Then, the logarithmic contribution to \eqref{eq:coll-sub-L-qe-alt} ensures that the upper limit on the $Q^2$ integral in the total  cross section is set by $Q^2_{\rm max}$, that is the explicit $\mu$ dependence cancels with that coming from \eqref{eq:xfgamma-phys}:
\be
 z p_{\gamma q}(z)  \left[\ln\frac{M^2(1-z)^2}{z\mu^2}F_2\left(\frac{x}{z},\mu^2\right) +   \int^{\frac{\mu^2}{1-z}}_{Q^2_{\rm min}} 
  \frac{dQ^2}{Q^2}F_2\left(\frac{x}{z},Q^2\right)\right] \to  z p_{\gamma q}(z)F_2\left(\frac{x}{z},\mu^2\right)\ln\frac{Q^2_{\rm max}}{Q^2_{\rm min}}\;.
\ee
The non--logarithmic in $Q^2$ contribution to the collinear cross section then reproduces, again under the above approximation, the contribution from the $M^2$ suppressed terms in \eqref{eq:sigma-HL-simple}:
\begin{align}\nonumber
 \int^{Q_\text{max}^2}_{Q_\text{min}^2}   \frac{dQ^2}{Q^2} \left[\frac{z ^2 Q^2}{M^2}
   -\frac{2 z  Q^2}{M^2}\right]F_2(x/z,Q^2) &\approx \int^{Q_\text{max}^2}_{Q_\text{min}^2}   \frac{dQ^2}{Q^2} \left[\frac{z ^2 Q^2}{M^2}
   -\frac{2 z  Q^2}{M^2}\right]F_2(x/z,\mu^2)\;,\\
   & \approx \left( 3z -2 - z^2\right)F_2(x/z,\mu^2)\;,
\end{align}
where the first line is true up to $O(\alpha_S, \alpha)$ corrections and the second line up to $O(Q^2/M^2)$ (and smaller $O(m_p^2/M^2)$) corrections. Subtracting the $-z^2$ term in \eqref{eq:xfgamma-MSbar} we arrive at the correct expression for $f^{NL}(z)$. This is no accident, as by applying the above approximations one is in effect isolating the corresponding LO, quark--initiated, contribution to $F_2$ in the complete expression and this must by construction match the corresponding expression calculated directly within the collinear approach. 

From the above discussion, one may in general expect a rather close matching between the complete structure function \eqref{eq:sigma-HL-simple} and collinear \eqref{eq:coll-sub-L-qe-alt} results. Certainly we can see that by including the NLO in $\alpha$ quark--initiated contribution, with the corresponding matching term in the photon PDF, this result will be closer to the structure function prediction. Nonetheless, the collinear result only reproduces the full structure function to a certain degree of approximation: in the collinear result the full $M^2$ dependence has been dropped, an artificial $\mu$ dependence (absent in the more precise structure function result) has been introduced and the complete result is only reproduced up to the  $O(\alpha_S, \alpha)$ corrections discussed above. 

\begin{figure}
\begin{center}
\includegraphics[scale=0.65]{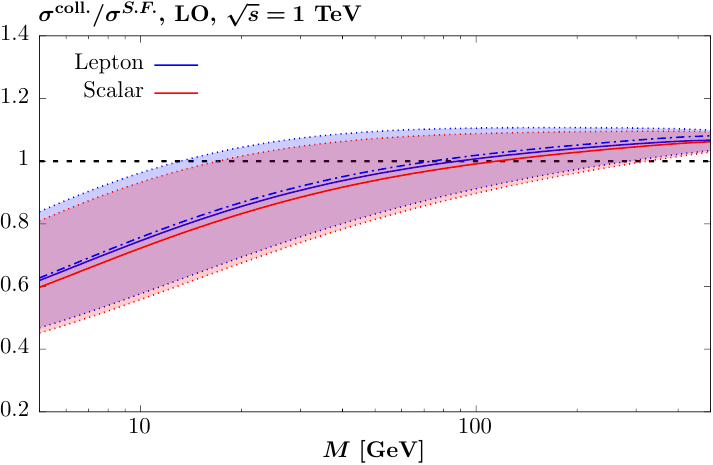}
\includegraphics[scale=0.65]{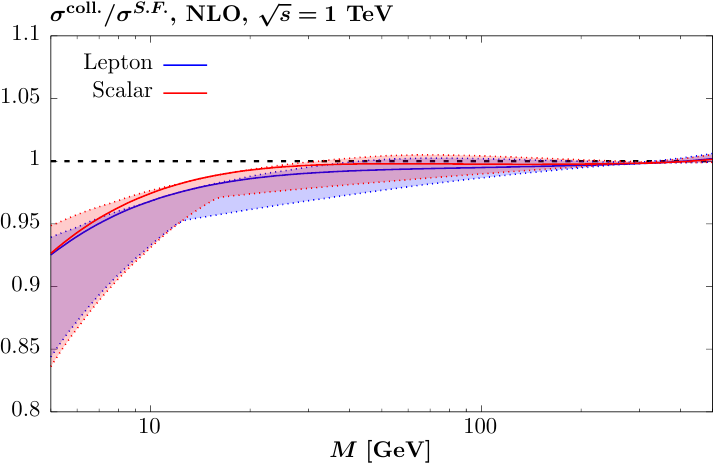}
\caption{Ratio of the cross sections for heavy lepton and scalar (with mass $M$) production calculated within the approximate collinear approach~\eqref{eq:coll-fact-basic}, to the structure function result~\eqref{eq:sigepa1}. The left (right) figures correspond to the LO $O(\alpha)$ (NLO $O(\alpha^2)$) cases, while in the LO case the result of the equivalent photon approximation is also shown for lepton production (dot--dashed line). The band correspond to variations in the factorization/renormalization scales by a factor of two around the central value $\mu=M$, indicated by the solid line.}
\label{fig:eparat}
\end{center}
\end{figure} 

To examine this effect further, in Fig.~\ref{fig:eparat} we show the ratio of the cross section calculated within the approximate collinear approach~\eqref{eq:coll-fact-basic}, to the more precise structure function result~\eqref{eq:sigepa1}, taking $\sqrt{s}=1$ TeV for concreteness.  For the collinear photon PDF we use the \texttt{MMHT2015qed\_nnlo} set~\cite{Harland-Lang:2019pla}, which is generated using a procedure that closely follows \texttt{LUXqed}, while for the structure function prediction we use the same inputs for the structure functions as in the collinear PDF. In particular, this is divided into elastic, inelastic resonance and inelastic continuum contributions in the usual way, and we have checked that our implementation can closely reproduce the collinear PDF. For the $Q^2 > 1$ ${\rm GeV}^2$ continuum component we use the ZM--VFNS at NNLO in QCD  predictions for the structure functions as implemented in~\texttt{APFEL}~\cite{Bertone:2013vaa}, with the \texttt{MMHT2015qed\_nnlo} PDFs. 

In general there is of course some uncertainty on the determination of the structure functions themselves, due principally to the uncertainty on their experimental determination, either directly or indirectly via the uncertainty on the quark/gluon PDFs which generate the photon at higher $Q^2$. In the case of the photon PDF, this is propagated through in~\cite{Harland-Lang:2019pla} to provide a photon PDF uncertainty, which is generally very small, entering at the percent level. In the structure function calculation, we must also in general include this source of uncertainty. However, we note that being of an almost identical origin, the uncertainty due to this will be highly correlated between the collinear and structure function calculations, and will therefore largely cancel in the ratio. Moreover, this source of uncertainty is of a fundamentally different origin to the effects, and in particular the scale variation uncertainty in the collinear calculation, that we wish to investigate here; these would be present even for perfect experimental knowledge of the structure functions. In the results in this section we therefore do not include any photon PDF uncertainty, or the corresponding uncertainty on the structure function calculation, though we will comment on these where relevant.


\begin{figure}
\begin{center}
\includegraphics[scale=0.65]{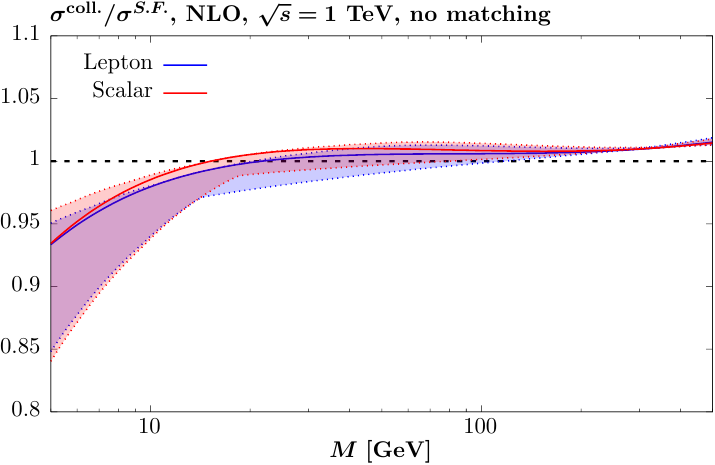}
\caption{As in Fig.~\ref{fig:eparat} (right), but excluding the `matching' term in~\eqref{eq:xfgamma-MSbar}.}
\label{fig:eparatnm}
\end{center}
\end{figure} 

For comparison, in addition to the case of heavy lepton production, we also consider the production of a heavy scalar $S$ via the $\gamma\gamma^* \to S$ subprocess, as discussed in~\cite{Manohar:2017eqh}. In this case we have
\begin{align}
\hat{\sigma}_T^S(\xi,Q^2) &=-\frac{2 \sigma_0}{s}\delta\left(\xi -x\right)\cdot (q\cdot k) \; ,\\
\hat{\sigma}_0^S(\xi,Q^2) &=0\;,
\end{align}
where $\sigma_0$ is an overall factor defined as in~\cite{Manohar:2017eqh} (with the missing factor of $\alpha$ in comparison to the lepton case due purely to the normalization convention defined there). Fig.~\ref{fig:eparat} (left) includes only the LO in $\alpha$ contribution to the collinear cross section, that is the first term in~\eqref{eq:coll-sub-L-qe-alt}, with the band given by the extrema found by varying the factorization and renormalization scales by the usual factor of 2 up and down around the default choice $\mu=M$. This gives a measure of the `uncertainty' one would evaluate when applying the collinear factorization approach, though it should be emphasised that this source of variation is completely absent in the structure function case. We can see that this variation can be rather large, being roughly $\sim \pm 30\%$ at lower mass, and approaching $\sim \pm 5-10\%$ at higher mass. Within this band the LO collinear prediction can differ by as much as a factor of $\sim 2$ at lower mass, while at highest mass it differs by $\sim 10\%$. At both low and high mass we note that the uncertainty does not cover the region from the structure function prediction. Thus we can see that the result of using the LO in $\alpha$ photon--initiated predictions, that is those directly proportional to the photon PDF, only produces the more precise structure function prediction rather approximately, in particular at lower masses. These differences are in particular significantly larger than the $\sim 1\%$ level PDF uncertainty on the photon PDF itself. We also show the lepton prediction excluding the `matching' term in~\eqref{eq:xfgamma-MSbar}, that is using the pure EPA result, and we can see that this has an extremely small impact relative to the effects discussed above. This is as expected, given that this term only enters formally at NLO in $\alpha$.

In Fig.~\ref{fig:eparat} (right) we show the same comparison as before, but now including the NLO quark--initiated contribution for the collinear cross section. We can see that as expected the impact of scale variation is smaller, and the agreement with the structure function result is much better. However, for lower masses the variation band is still $\sim \pm 5\%$, and even in the intermediate mass region is of the same $\sim 1\%$ level  as the photon PDF uncertainty. At high mass, the agreement is very good and the scale variation is at the  $\sim 0.5\%$ level or smaller. For the sake of comparison, we also show in Fig.~\ref{fig:eparatnm} the same results, but now excluding  the `matching' term in~\eqref{eq:xfgamma-MSbar}, and we can see that this indeed results in a slightly larger variation band, with some $\sim 1\%$ level  disagreement with the structure function result, not covered by the variation, at the highest masses.

\begin{figure}
\begin{center}
\includegraphics[scale=0.65]{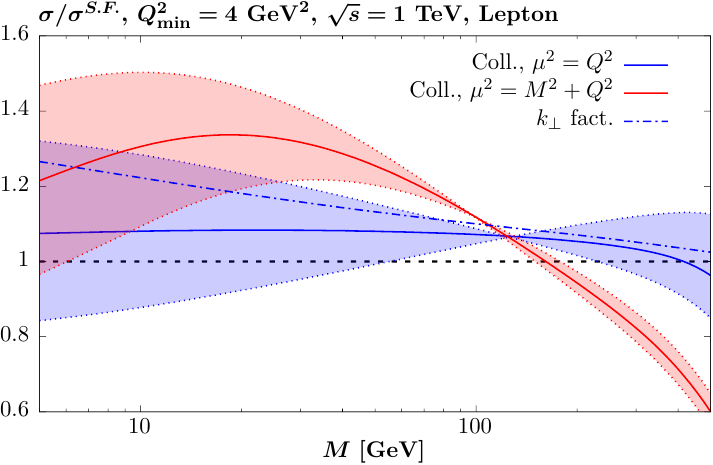}
\includegraphics[scale=0.65]{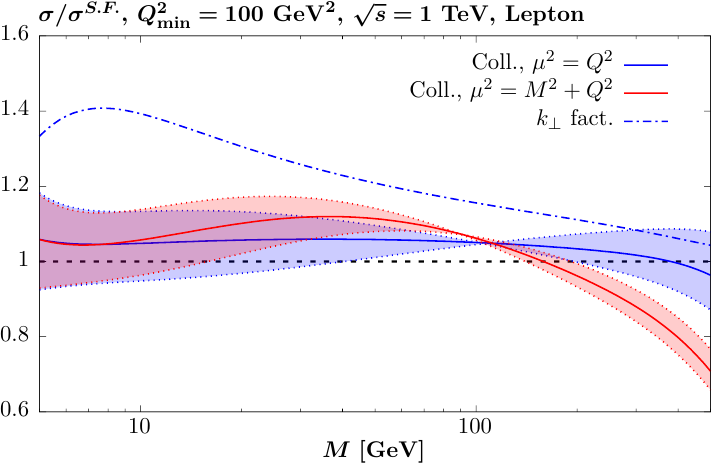}
\caption{Ratio of the cross sections for heavy lepton (with mass $M$) production calculated within the approximate NLO collinear approach~\eqref{eq:coll-fact-basic}, to the structure function result~\eqref{eq:sigepa1}. The left (right) figures correspond to the case that a cut of $Q^2 > $ 4 (100) ${\rm GeV}^2$ is placed. Results for different choices of factorization scale are shown, with the band corresponding to variations in the factorization/renormalization scales as above. The prediction within the $k_\perp$ factorization approach is also shown.}
\label{fig:eparat_diff}
\end{center}
\end{figure}

Having considered the inclusive cross section for heavy lepton/scalar production, we next introduce a cut into the analysis, namely by requiring that the photon $Q^2$ be larger than a particular value. In this case the LO collinear photon--initiated channel gives zero contribution and therefore the first non--zero collinear contribution enters at NLO. The corresponding subprocess cross sections are
\begin{align}
\frac{{\rm d}\hat{\sigma}^{L}}{{\rm d} Q^2} &=c_0 \cdot \frac{\alpha^2(Q^2) }{2 \pi Q^2}e_q^2 \left( z p_{\gamma q}(z) + z(z-2)\frac{Q^2}{M^2}\right)\;,\\
\frac{{\rm d}\hat{\sigma}^{S}}{{\rm d} Q^2} &=\sigma_0 \cdot \frac{\alpha(Q^2) }{2 \pi Q^2} e_q^2\left(z p_{\gamma q}(z)  + 2\frac{Q^2}{M^2}z(z-1) + \frac{Q^4}{M^4}z^2 \right)\;,
\end{align}
which are then folded with~\eqref{eq:coll-fact-basic} in the usual way. These results can essentially be read off from the full structure function expressions, by substituting the LO in QCD expressions for the structure functions and working in the massless proton limit.

An additional possibility for modelling these processes at non--zero photon $Q^2$, or equivalently transverse momentum, is to apply the $k_\perp$--factorization approach. Here, we can use~\eqref{eq:xfgamma-phys} to define an unintegrated PDF, by keeping the $Q^2$ (or equivalently, the photon transverse momentum, $k_\perp$) dependence unintegrated over, see~\cite{daSilveira:2014jla,Luszczak:2015aoa,Dyndal:2019ylt,Manohar:2017eqh} for further discussion. The subprocess cross section is extracted from the usual approximate expression for the off--shell photon density matrix applied within this approach:
\be
\frac{2k_{\perp}^\mu k_\perp^\nu}{k_\perp^2} L_{\mu\nu} \approx \delta_T^{\mu\nu}L_{\mu\nu} \;,
\ee
where the second approximate equality is true in the $Q^2 \ll M^2$ limit and the transverse projection is as defined in \eqref{eq:projdef}. This allows us to write
\be
\frac{{\rm d} \sigma^{k_\perp {\rm fact.}}}{{\rm d} Q^2} =    \int {\rm d} \xi\, \hat{\sigma}_T(\xi) \cdot \frac{{\rm d }f_{\gamma/p}^{\rm PF}(\xi,Q^2)}{{\rm d}Q^2}\;.
\ee
This therefore gives a prediction for the $Q^2$ dependence of the cross section at LO, unfolding the integration that is implicit in the usual photon PDF, but we can immediately see that it will miss the full kinematic dependence of the more precise structure function cross section, both due to the missing contribution from longitudinal photon polarizations as well as due to the $M^2 \gg Q^2$ approximation made for the transverse.

\begin{figure}
\begin{center}
\includegraphics[scale=0.63]{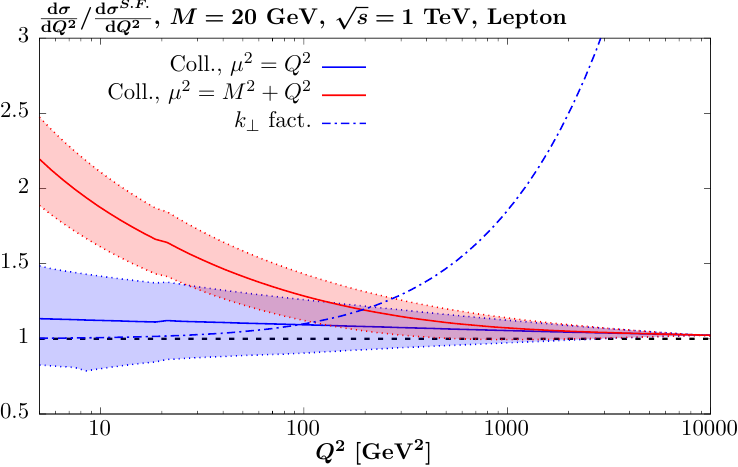}
\includegraphics[scale=0.63]{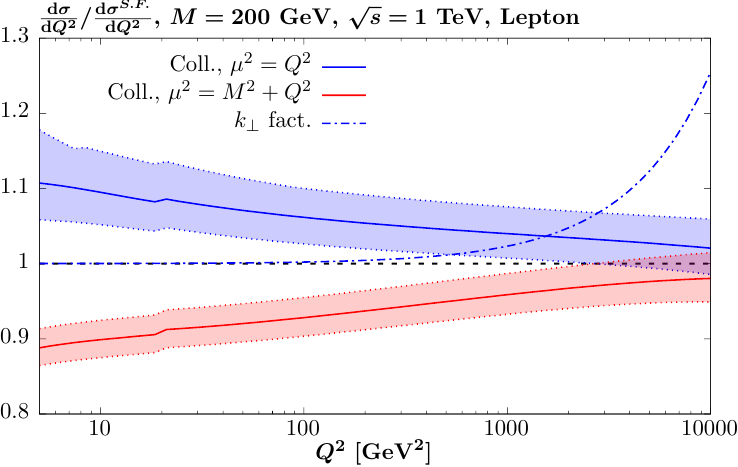}
\caption{Ratio of the differential cross sections with respect to the photon virtuality, $Q^2$, for heavy lepton  (with mass $M$) production calculated within the approximate collinear approach~\eqref{eq:coll-fact-basic}, to the structure function result~\eqref{eq:sigepa1}. Results for different choices of factorization scale are shown, with the band corresponding to variations in the factorization/renormalization scales as above. The prediction within the $k_\perp$ factorization approach is also shown. The two figures correspond to different choices of  lepton mass, as indicated.}
\label{fig:eparat_diffq}
\end{center}
\end{figure} 

The results are shown in Fig.~\ref{fig:eparat_diff}, where for clarity we only show the case of lepton production, finding the scalar case to be rather similar. For the collinear prediction we consider two choices for the factorization scale of the quark PDFs, namely $\mu_F^2= Q^2$ and  $\mu_F^2= Q^2+M^2$. While the former may be considered a more natural and sensible choice, as in the structure function calculation the cross section is written in terms of structure functions evaluated at scale $Q^2$, the latter is the type of scale one might be tempted to take from the point of view of a `standard' QCD calculation. In particular, for the $Z$ boson $p_\perp$ distribution we will consider in the following section, a standard choice would be $\mu_F^2 = (p_\perp^Z)^2 + M_Z^2$, which is analogous to this. 

The $Q^2 > 4$ ${\rm GeV}^2$ case is shown in Fig.~\ref{fig:eparat_diff} (left). For $\mu_F^2= Q^2$ we can see that the discrepancy with the structure function result from the collinear prediction, within the scale variation band, is as large as $\sim \pm 20 \%$, with differences of this order continuing up to high mass. Indeed, in certain regions the difference is not covered by the scale variation band itself. This level of difference is entirely consistent with the missing higher order corrections that are omitted in the explicit collinear calculation but which are fully accounted for in the more precise structure function result. The prediction within $k_\perp$--factorization differs by a similar amount, while for the $\mu_F^2= Q^2+M^2$ choice this effect is even more severe. In Fig.~\ref{fig:eparat_diff} (right) we impose a higher $Q^2 > 100$ ${\rm GeV}^2$ cut, and find the difference and scale variation band for the collinear predictions are somewhat smaller, though still not negligible, while for the $k_\perp$--factorization case the difference is larger, as one would expect from the fact that this assumes $Q^2 \ll M^2$ to be true.

To clarify things further, in Fig.~\ref{fig:eparat_diffq} we show the differential cross sections, for two choices of lepton mass, $M=20$ and 200 GeV. Similar levels of difference can be seen as above, particularly at lower $Q^2$ for the collinear predictions, while for $k_\perp$--factorization the approach is seen to break down at large $Q^2 \gtrsim M^2$, as one would expect. As before, the scale variation bands do not necessarily cover the more precise structure function result.

In summary, we have seen that for the inclusive cross section, provided only the LO in $\alpha$ photon--initiated contribution is included, the uncertainty on the corresponding predictions is significantly larger than that coming from the very small quoted PDF uncertainty, and at low masses the standard scale variation band does not overlap with the more precise structure function result. This is of particular significance to LHC phenomenology, where the inclusion of photon--initiated production at LO is common. The situation improves to a large extent when the NLO collinear contribution is included, although even here the uncertainty at lower mass is again significantly larger than the corresponding PDF uncertainty and even at higher masses of the same order. However, such corrections are not always available (publicly or otherwise) for LHC processes. Moreover, even if these corrections are eventually explicitly included, one will still introduce an (albeit smaller) source of uncertainty due to the residual scale dependence that can be bypassed entirely by simply working with the more precise structure function result, as calculated in the structure function approach. More significantly from a phenomenological point of view, we have seen that once one starts to include cuts, or consider observables that are sensitive to the photon transverse momenta, the difference between even the NLO prediction (or that using the $k_\perp$--factorization approach) can again be rather large.

We note that the magnitude of these scale variation uncertainties in the inclusive cross sections are roughly consistent with the LO and NLO uncertainty bands on the photon PDF presented in Section 9 of~\cite{Manohar:2017eqh}, being of a similar origin. However, here the final `missing higher order' uncertainty derived within this approach is, as discussed in this work (see footnote 11), only relevant for the case that one works at NLO for the photon--initiated contributions, and will otherwise drastically underestimate the corresponding uncertainty, as we have seen above. Moreover even if one works at NLO, then the uncertainty that they include, which comes from the manner in which one defines the photon PDF and the factorization scale choice which corresponds to it, is entirely absent in the structure function calculation. More significantly, while this uncertainty is estimated to be rather small in~\cite{Manohar:2017eqh}, at the $\sim 1\%$ level or less, the scale variation uncertainty in the NLO collinear cross section is not entirely accounted for by this, and is in many cases larger, as we have seen. On the other hand, as discussed at the end of Section~\ref{sec:structfun}, other small sources of uncertainty from missing higher--order non--factorizable corrections, remain in both the structure function and collinear calculations.

\section{Hadron--hadron collisions}\label{sec:hadhad}

We now consider some phenomenological implications of the results above for photon--initiated production at the LHC. Before doing so, we briefly discuss the connection between the structure function result~\eqref{eq:sighh} and the collinear prediction via the photon PDF, similarly to the lepton--hadron case considered before. As in~\cite{Harland-Lang:2019zur} we can write
 \be\label{eq:sighhf}
 \sigma_{pp} = \frac{1}{2s}  \int  {\rm d}x_1 {\rm d}x_2\,{\rm d}^2 q_{1_\perp}{\rm d}^2 q_{2_\perp
} {\rm d \Gamma} \,\alpha(Q_1^2)\alpha(Q_2^2) \frac{\rho_1^{\mu\mu'}\rho_2^{\nu\nu'} M^*_{\mu'\nu'}M_{\mu\nu}}{q_1^2q_2^2}\delta^{(4)}(q_1+q_2 - p_X)\;,
 \ee
 where $x_i$ and $q_{i\perp}$ are the photon momentum fractions (see~\cite{Harland-Lang:2019zur} for precise definitions) and transverse momenta, respectively.
The amplitude squared $M^*_{\mu'\nu'}M_{\mu\nu}$ permits a general expansion~\cite{Budnev:1974de}
\be
M^*_{\mu'\nu'}M_{\mu\nu} = R_{\mu\mu'} R_{\nu\nu'}  \,\frac{1}{4}\sum_{\lambda_1 \lambda_2} |M_{\lambda_1 \lambda_2}|^2 + \cdots\;,
\ee
where we omit various terms that vanish when taking the $Q_{1,2} \ll M_X^2$ limit, or after integration over the photon azimuthal angle. Here $R$ is the metric tensor that is transverse to the photon momenta $q_{1,2}$:
\be
R^{\mu\nu} = -g^{\mu\nu} + \frac{(q_1 q_2)(q_1^\mu q_2^\nu+q_1^\nu q_2^\mu)+Q_1^2 q_2^\mu q_2^\nu+Q_2^2 q_1^\mu q_1^\nu}{(q_1 q_2)^2-Q_1^2 Q_2^2}\;.
\ee
We are then interested in
\begin{align}\nonumber
\rho_{\mu\nu}^1 R^{\mu\nu} &= 2\int \frac{{\rm d} x_{B_1}}{x_{B_1}} \frac{1}{x^2_{B_1}} \bigg[\left(\frac{1}{2}\left\{1+\frac{((q_1 q_2)-2 x_{B_1} (q_2 p_1))^2}{(q_1 q_2)^2-Q_1^2 Q_2^2}\right\}+\frac{2 m_p^2 x_{B_1}^2}{Q_1^2}\right)F_2(x_{B_1},Q^2_1)\\
&-F_L(x_{B_1},Q^2_1)\bigg]\;,
\end{align}
and similarly for $\rho^2$, after interchanging $1\leftrightarrow 2$. Here the change of variables from ${\rm d} M_i^2$ to ${\rm d}x_{B_i}$ is permitted as we now integrate with respect to $Q^2_i$, as described below. Then dropping the subleading $Q_{1,2}^2$ terms and using that $(q_1 q_2) \approx x_1 (p_1 q_2)$ in this limit, we get
\be
\rho_{\mu\nu}^i R^{\mu\nu} \approx 2\int \frac{{\rm d} z_i}{z_i} \frac{1}{x^2_{i}} \left[\left(z_i p_{\gamma q}(z_i)+\frac{2x_i^2 m_p^2}{Q_i^2}\right)F_2(x_i/z_i,Q^2_i)-z_i^2 F_L(x_i/z_i,Q^2_i)\right]\;,
\ee
where $z_i=x_i/x_{B_i}$ as usual, and $i={1,2}$. Performing the angular integration and replacing\footnote{Strictly speaking, this introduces a factor of $(1-x_1)(1-x_2)$, but as discussed in~\cite{Harland-Lang:2019zur} this cancels with the corresponding kinematic factor that is present when one moves away from the high energy limit.} ${\rm d}q_\perp^2 = {\rm d} Q^2$, we readily find that
\be
\sigma_{pp} \approx \int  {\rm d}x_1 {\rm d}x_2\,f_{\gamma/p}^{\rm PF}(x_1,\mu^2)f_{\gamma/p}^{\rm PF}(x_2,\mu^2)\hat{\sigma}(\gamma\gamma \to X)\;,
\ee
where we have used $x_1 x_2 s\approx M_X^2$, and the usual form for the photon--initiated cross section in terms of the squared matrix element and phase space measure. As expected, the cross section is approximately given in terms of the same `physical' photon PDFs as in \eqref{eq:xfgamma-phys}, with a `factorization' scale that in this case again reflects the lack of control over the $Q_{1,2}^2 \sim M_X^2$ region in this approximation.

\begin{figure}
\begin{center}
\includegraphics[scale=0.64]{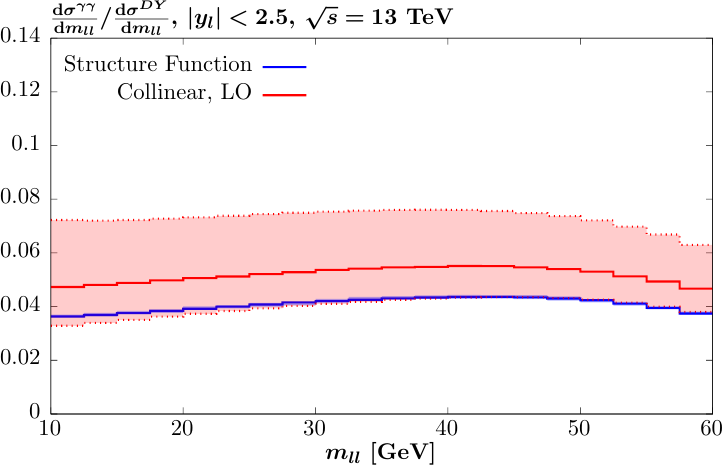}
\includegraphics[scale=0.64]{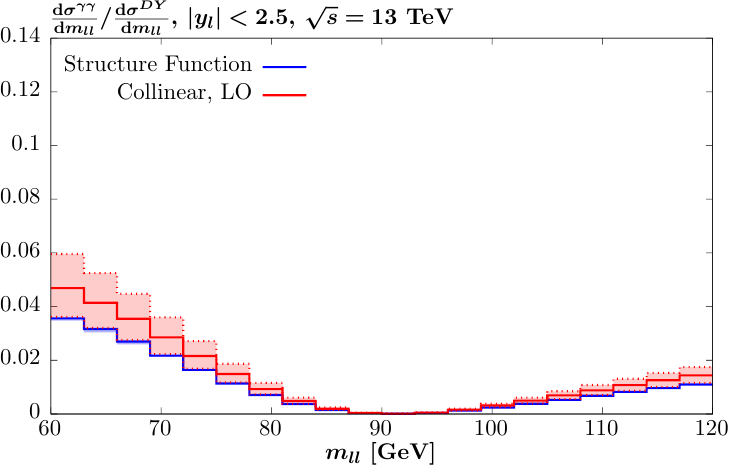}
\includegraphics[scale=0.64]{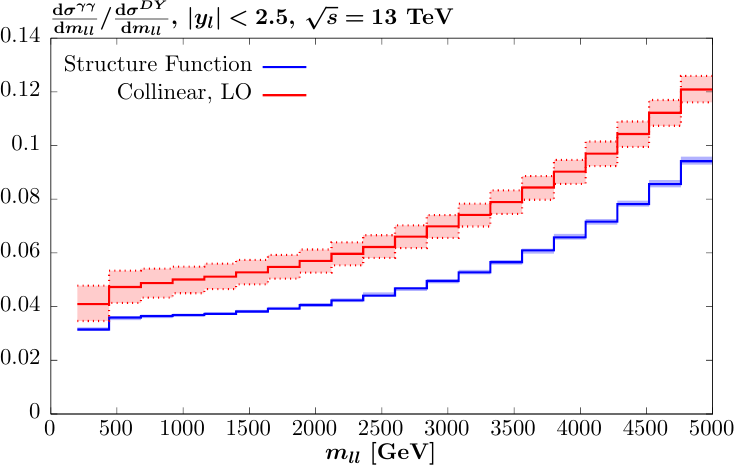}
\caption{Ratio of the photon--initiated cross sections for lepton pair production production to the NLO QCD Drell--Yan cross section at the $13$ TeV LHC, as a function of the lepton pair invariant mass, $m_{ll}$. The LO collinear predictions and the more precise structure function result, using~\eqref{eq:sighh} directly, are shown. In the former case the uncertainty band due to factorization scale variation by a factor of two around the central value $\mu=m_{ll}$, is given. The leptons are required to lie in the $|\eta_l|<2.5$ region. In the latter case, the error band due to the experimental uncertainty on the on the structure functions is shown (note in some regions this is comparable to the line width of the central value).}
\label{fig:mDY}
\end{center}
\end{figure}

\begin{figure}
\begin{center}
\includegraphics[scale=0.655]{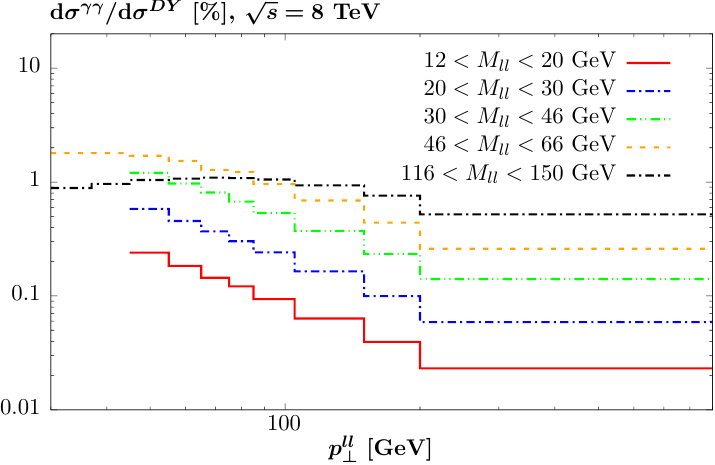}
\includegraphics[scale=0.655]{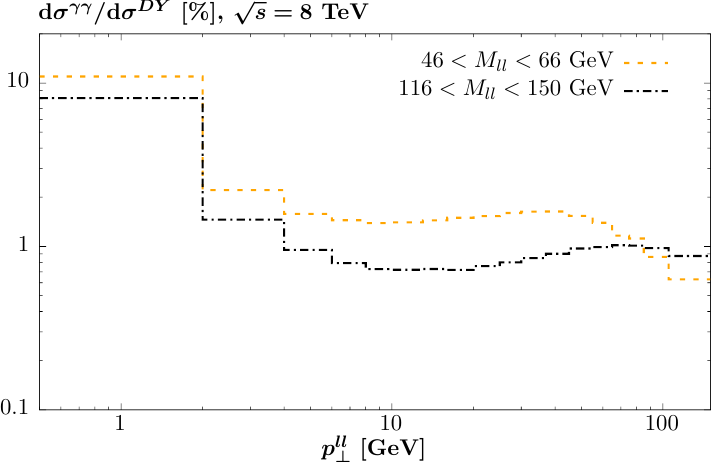}
\caption{Percentage contribution from photon--initiated production to the lepton pair $p_\perp$ distribution, within the ATLAS~\cite{Aad:2015auj} off--peak event selection, at 8 TeV. The photon--initiated cross section is calculated using~\eqref{eq:sighh} directly, while the QCD predictions in the left (right) plots correspond to NNLO (NNLO+NNLL) QCD theory.}
\label{fig:theorypt}
\end{center}
\end{figure}

We now consider some phenomenological applications of the structure function approach. In Fig.~\ref{fig:mDY} we show the fraction of the photon--initiated contribution to the Drell--Yan production of lepton pairs at the 13 TeV LHC. For the photon--initiated prediction we show the LO collinear results, given in terms of the \texttt{MMHT2015qed\_nnlo} photon PDF, with the uncertainty band due to factorization/renormalization scale variation by a factor of two around the central value $\mu=m_{ll}$, shown, assuming $\mu_R = \mu_F$. The fixed--order QCD predictions are made at NLO using \texttt{APPLgrid}~\cite{Carli:2010rw} + \texttt{MCFM}~\cite{Boughezal:2016wmq}. We also plot the structure function result found by using~\eqref{eq:sighh} directly, within the structure function approach. A rather larger scale variation band is evident, in particular at lower masses where it is $\sim 50\%$. The more precise structure function results tend to lie on the lower end of the variation band, while at high mass it in fact lies outside the uncertainty suggested by scale variation. In the latter case, this may be because the more appropriate factorization scale contains some $z$ dependence, in order to reproduce the impact of the correct kinematic limit on the $Q^2$ integration (see the discussion below \eqref{eq:xfgamma-phys}), with this effect becoming more significant in the high mass ($z \to 1$) region.

For the structure function calculation, we also show the uncertainty due to the experimental inputs on the structure functions. These are evaluated following the procedure discussed in~\cite{Harland-Lang:2019pla}, which is closely based on that described in~\cite{Manohar:2016nzj,Manohar:2017eqh}. We refer the reader to these references for further details, but in summary we include: an uncertainty on the A1 collaboration~\cite{Bernauer:2013tpr} fit to the elastic proton form factors, based on adding in quadrature the experimental uncertainty on the polarized extraction and the difference between the unpolarized and polarized; a $\pm 50\%$ variation on the ratio $R_{L/T}$, relevant to the low $Q^2$ continuum inelastic region; a variation of $W^2_{\rm cut}$, the scale below which we use the CLAS~\cite{Osipenko:2003bu} fit to the resonant region, and above which we use the HERMES~\cite{Airapetian:2011nu} fit/pQCD calculation (for $Q^2$ below/above $1\,{\rm GeV}^2$), between $3$--$4$ GeV; the symmetrised difference between the default CLAS and Cristy--Bosted~\cite{Christy:2007ve} fits to the resonant region; the standard PDF uncertainty on the \texttt{MMHT2015qed\_nnlo} quark and gluon partons. For simplicity we do not include a renormalon correction or associated uncertainty, as this would require modification of the \texttt{APFEL} evolution code, but note that this is a small effect. 

We can see that the uncertainty on the structure function calculations due to the input structure functions is very small, at the percent level, and often barely visible on the plots. This is consistent with our expectations based on the similarly small quoted uncertainties on the photon PDF, which are of an almost identical origin. For the sake of clarity we do not show the PDF uncertainty on the collinear result, which we have checked is indeed in general comparable to the above uncertainty on the structure function calculation, and in particular much smaller than the scale variation uncertainty, and in the case of high mass production, the difference between the structure function and LO collinear results.

In addition to providing a more precise prediction for inclusive lepton pair production at the LHC, we can apply the structure function approach to make precision predictions for a process that cannot be evaluated at all using the LO collinear photon--initiated calculation, namely the lepton pair transverse momentum distribution\footnote{The contribution from initial--state $Z$ bosons, which may play a role in particular at larger $p_\perp^{ll}$, is not included here. In addition, in these figures we do not include those diagrams where only one photon couples to the lepton pair, see e.g. Figs. 4 (a) and (d) of~\cite{CarloniCalame:2007cd}, which may play some role in the $p_\perp^{ll} \gtrsim m_{ll}$ region, but are otherwise kinematically suppressed. We leave a full inclusion and discussion of these contributions, both of which can be readily accounted for within the structure function approach, to future work.}. Here, the cross section is zero at LO within the collinear approach. We in particular consider the ATLAS 8 TeV measurement~\cite{Aad:2015auj}, which has the advantage of being presented both on and off the $Z$ peak.  While in the former case we expect the contribution to be negligible, as we have indeed checked, in the latter case this may not be true. The leptons are required to have $p_\perp^l >20$ GeV and $|\eta_l|<2.4$. The percentage photon--initiated contributions, with respect to the predicted QCD DY cross section, are shown in Fig.~\ref{fig:theorypt}. In the left plot results for the $p_\perp^{ll}>30 $ GeV region, where fixed--order QCD predictions for the DY process may be trusted, are shown. The QCD predictions are calculated as in the previous case, but now supplemented with NNLO K--factors produced with \texttt{NNLOjet}~\cite{Bizon:2018foh}. We can see that the photon--initiated contributions are small, but not necessarily negligible, being at the percent level at lower  $p_\perp^{ll}$. There is a clear trend for the relative contribution to decrease both with decreasing $m_{ll}$ (note that the cuts applied are rather different from that shown in Fig.~\ref{fig:mDY}) and and with increasing $p_\perp^{ll}$. In the right plot we show predictions for the low $p_\perp^{ll}$ region, comparing to NNLO+N${}^3$LL resummed predictions produced with \texttt{NNLOjet+RadISH}~\cite{Bizon:2018foh}. These are again found to be small but not negligible, in particular in the larger invariant mass bin. Moreover, in the lowest $p_\perp^{ll}$ bin the contribution from photon--initiated production, which will contain essentially all of the elastic contribution as well as the majority of the resonant and low--mass continuum inelastic, is large, being $\sim 10\%$.  This is explained by the fact that while the QCD contribution in this region is strongly Sudakov suppressed, for the photon--initiated contribution no such effect is present and indeed the cross section is peaked (though by construction finite) in this region. 
 
 \section{Summary and Outlook}\label{sec:conc}
 
In light of the high precision LHC physics programme that lies ahead of us, a precise account of photon--initiated channels in proton--proton collisions is vital. In this paper we have critically re--examined the current state of the art for the calculation of such processes, which is formulated in terms of a photon PDF in the proton that may be determined rather precisely from the known proton structure functions. We have in particular demonstrated how a rather straightforward application of the well--known structure function approach provides the most precise available calculation of photon--initiated production. This approach, which is well known from the case of Higgs boson production via VBF, is by construction more precise than a direct calculation with the photon PDF, and indeed bypasses any reference to this object entirely. In particular, we have shown how the calculation in terms of a photon PDF is derived from an approximate form of the full structure function calculation. 

The calculation in terms of collinear factorization and the photon PDF in particular introduces a factorization scale variation in the predicted cross section that is purely an artefact of this particular formulation, and which is absent in the structure function prediction. This is particularly relevant in light of the fact that in many cases only the LO in $\alpha$ photon--initiated matrix elements are used; as we have studied in detail in this paper, these suffer from rather large scale variation uncertainties. While these will naturally improve when the NLO corrections are included, we have nonetheless seen that the residual variation at NLO is not negligible in comparison to the quoted photon PDF uncertainties, particularly so for more differential observables and/or when cuts are applied.
 
 With this in mind, we have presented high precision predictions for lepton pair production at the LHC, via the structure function approach. This is in contrast to the current standard collinear predictions, which are applied at LO in $\alpha$ and hence suffer from rather large scale variation uncertainties at up to the $\sim 50\%$ level. We have in addition presented the first precision predictions for the photon--initiated contributions to the lepton pair transverse momentum distribution, within the ATLAS 8 TeV event selection; such a contribution is zero within the collinear approach at LO. These are found to be small, but not negligible, entering at the percent level in some regions of phase space. These contribute in the kinematic regions relevant to both the fixed order pQCD calculation of the Drell--Yan process and the region of lower transverse momenta, where resummation must be applied. The structure function approach allows the contribution in both regions to be calculated consistently and straightforwardly.
 
 What do the results of this paper imply for the inclusion of QED effects at the LHC? For the calculation of the photon--initiated contributions, we have shown that the structure function approach provides the most precise available calculation, and therefore should be used. From a practical point of view, the application of this approach is rather straightforward, though the dependence of such predictions on the input quark/gluon PDFs is currently not included in the standard~\texttt{ApplGrid}~\cite{Carli:2010rw} and~\texttt{Fastnlo}~\cite{Wobisch:2011ij} tools; photon--initiated production is included in the former case but only in terms of an explicit photon PDF. The extension of these to include predictions within the structure function approach is however certainly possible and desirable. In particular, the dependence of the hadronic cross section on the input quark/gluon PDFs, which enters through the calculation of the high $Q^2$ contribution to the structure functions, is perfectly amenable to such tools. From the point of view of global PDF fits, it would be useful to provide publicly available grids for the structure functions, which could then be straightforwardly used for LHC phenomenology; this is something the MMHT collaboration plan to do in future releases.

 The inclusion of QED corrections to the DGLAP evolution can proceed in the same way as discussed in~\cite{Harland-Lang:2019pla,Bertone:2017bme}, that is via an input photon PDF defined through the LUXqed procedure, which is itself based implicitly on the structure function approach. The calculation of~\cite{Harland-Lang:2019pla}, which explicitly evolves a photon PDF defined at low scale $Q_0$ within such an approach, provides a particularly natural application of this. A final complication that may arise is the systematic account of collinear $\gamma \to q\overline{q}$ splittings, if these may contribute to a particular process, when the initiating photon is included within the structure function approach. It should however be possible to account for these within the structure function approach, by simply subtracting the contribution from the DGLAP evolution of the corresponding quark PDF at the fixed order at which one is calculating, to avoid any double counting.

 In this paper we have considered lepton pair production at the LHC, as a natural first candidate, but clearly this is not the only process of relevance. Further natural extensions would include such processes as $W$ pair and Higgs production (for which in~\cite{Ciccolini:2007ec} percent level contributions are found, albeit with a rather outdated photon PDF set). These will be released as part of the public \texttt{SuperChic} MC~\cite{Harland-Lang:2018iur} implementation, for both the inclusive case considered here and the semi--exclusive case, that is with rapidity gaps in the final--state. In the meantime, predictions for lepton pair production, for arbitrary kinematics, are available upon request.

\section*{Acknowledgments.}

I thank Valery Khoze, Luca Rottoli, Misha Ryskin, Robert Thorne and Alessandro Vicini for useful discussions, Valerio Bertone for guidance in the use of \texttt{APFEL} and Alexander Huss for providing NNLO and NNLO+N${}^3$LL QCD predictions for the $Z$ boson $p_\perp$ distributions. I thank the Science and Technology Facilities Council (STFC) for support via grant award ST/L000377/1.

\bibliography{references}{}
\bibliographystyle{h-physrev}

\end{document}